\documentclass[aps,pra,twocolumn]{revtex4-1}

\usepackage{amsbsy,latexsym,amsmath}
\usepackage{amsfonts}
\usepackage{amssymb}
\usepackage[mathscr]{eucal}
\usepackage{epsfig,graphics,graphicx}
\usepackage{xcolor}
\usepackage{hyperref}
\hypersetup{colorlinks=true,citecolor=blue}

\newcommand{\ket}[1]{\left|{#1}\right\rangle}
\newcommand{\bra}[1]{\left\langle{#1}\right|}
\newcommand{\braket}[2]{\langle{#1}|{#2}\rangle}
\newcommand{\ketbrad}[1]{\left|{#1}\rangle\!\langle{#1}\right|}

\renewcommand{\P}{\mathbf{P}_n}
\renewcommand{\r}{\mathbf{r}}

\begin{document}

\title{Online strategies for exactly identifying a quantum change point}
\author{Gael Sent\'{\i}s$^{1}$}\email{gael.sentis@uni-siegen.de}
\author{Esteban Mart\'{\i}nez-Vargas$^{2}$}\email{esteban.martinez@uab.cat}
\author{Ramon Mu\~{n}oz-Tapia$^{2}$}\email{ramon.munoz@uab.cat}
\affiliation{
$^{1}$Naturwissenschaftlich-Technische Fakult\"at, Universit\"at Siegen, 57068 Siegen, Germany\\
$^{2}$F\'{i}sica Te\`{o}rica: Informaci\'{o} i Fen\`{o}mens Qu\`antics, Departament de F\'{\i}sica, Universitat Aut\`{o}noma de Barcelona, 08193 Bellaterra (Barcelona), Spain}

\begin{abstract}
We consider online detection strategies for identifying a change point in a stream of quantum particles allegedly prepared in identical states. We show that the identification of the change point can be done without error via sequential local measurements while attaining the optimal performance bound set by quantum mechanics. In this way, we establish the task of exactly identifying a quantum change point as an instance where local protocols are as powerful as global ones. The optimal online detection strategy requires only one bit of memory between subsequent measurements, and it is amenable to experimental realization with current technology.

\end{abstract}
\maketitle

\section{Introduction}

The ability to process streaming data on-the-fly and promptly detect changes in trends has become a most desirable feature of modern data-analysis algorithms. 
Change point detection is a vast branch of statistical analysis~\cite{brodsky,baseville} devoted to techniques for uncovering abrupt changes in the underlying probability distribution that generates a stream of stochastic data. Applications are far-reaching, including quality control~\cite{Lai1995}, medical diagnosis~\cite{Rosenfield2010}, and robotics~\cite{robots}. Generically, there are two distinct approaches for detecting change points: \emph{offline} strategies that require availability of a complete time series of data, and \emph{online} strategies that are able to process data sequentially. Naturally, having access to the full data history of a given stochastic process typically results in higher change-point identification rates. On the other hand, online strategies enable real-time decision making, are more versatile, and require less memory. These are most relevant in machine learning, for devising online algorithms with effective mechanisms to address learning in the context of non-stationary distributions, a problem known as concept drift~\cite{Webb2016a}.

The first extension of the change point identification problem, in its simplest formulation, into a quantum setup was recently introduced in Refs. \cite{cp-1,cp-2}. 
The problem can be stated as follows.
A source assumed to prepare a sequence of quantum particles in identical states suffers a sudden alteration at some unspecified point, after which the particles are prepared in a mutated state.
Given a sequence of particles, one aims at  
detecting when the mutation took place. 
In the most fundamental setting, that we also consider here, the initial and final states are assumed to be pure and known and, for a given sequence of length $n$,
all potential positions of the change point in the sequence are expected to happen with equal probability.
In Ref.~\cite{cp-1}, the minimum probability of erroneously identifying a quantum change point and a strategy that achieves it were obtained. This optimal strategy consists in a quantum measurement acting coherently on the given sequence of $n$ particles. It was also shown that a fairly general class of online strategies, based on sequential adaptive measurements on each individual quantum particle, underperform the optimal protocol, and strong numerical evidence that this is the case for all online strategies was provided. 
The experimental implementation of adaptive online strategies for change point detection has been very recently demonstrated~\cite{exp-cp}.
In contrast, Ref.~\cite{cp-2} addressed the quantum change point problem from a different approach: when no identification errors are allowed. The identification protocol then has two possible outcomes, either a correct answer or an inconclusive one~\cite{unambiguous}, and optimality means achieving a minimal rate of inconclusive outcomes. This scenario covers situations where, after the identification of a change point, a response action shall be taken only in conditions of absolute certainty.
The optimal procedure and its associated optimal success probability were derived analytically for any length $n$ and arbitrary states~\cite{cp-2}. Again, this optimal protocol would in principle require a coherent quantum measurement over the full sequence of particles, and hence also quantum memories to store them, which may render the protocol impractical in some scenarios.
In this paper, we look into online strategies for  
exactly identifying a change point in streaming quantum data.
Some simple online protocols were already considered in Ref.~\cite{cp-2} and shown to significantly underperform the optimal global protocol. Here, we deepen the analysis and address more general online strategies by allowing classical communication between local measurements. Contrary to our initial conjecture~\cite{cp-2}, we find the striking result that there is an online strategy that does achieve optimal performance up to a critical value of the overlap between the reference and mutated state. We also obtain that 
only one bit of 
memory is
required at each measurement step
to achieve optimality:
it is enough to know whether
the previous result was inconclusive. 
Our results hence imply that the exact optimal identification of a quantum change point is a readily implementable task with current technology, thus prone to integration within diverse quantum information processing protocols.

We begin by setting the notation and briefly reviewing the results for the optimal (global) strategy in Section~\ref{sec:global}. Then, we turn to online strategies in Section~\ref{sec:local} and present our core results. We show that these provide optimal performance in a given range of the overlap parameter. 
Beyond this range, we show that the best online strategy is, albeit suboptimal, very close to optimality. We finish in Section~\ref{sec:conclusions} with a short discussion.

\section{Optimal global strategy}\label{sec:global}
Let us denote by $\ket{0}$ the default state,  $\ket\phi$ the mutated state, and $c=\braket{0}{\phi}$ their overlap.
Without loss of generality, we take $c$ real and non-negative.
Given a sequence of $n$ particles, the change point identification corresponds to identifying a state within the set of equally likely source states %
$ \{\ket{\Psi_k}\}_{k=1}^n$, where
\begin{equation}
\label{psik}
\ket{\Psi_k}= |\underbrace{0\ldots0}_{k-1}\underbrace{\phi \ldots\phi}_{n-k+1}\rangle
\end{equation}
is associated with the change point occurring at position $k$. 
A strategy that unambiguously identifies the correct source state is characterized by a positive operator valued measure (POVM) with $n+1$ elements $\{E_l\geq 0\}_{l=0}^n$.
The outcomes $l=1,\ldots,n$ detect without error each possible source state, i.e., the corresponding POVM elements fulfill $\bra{\Psi_k}E_l\ket{\Psi_k}=0$ for $k\neq l$, and the remaining element $E_0=\openone -\sum_{k=1}^{n}E_{k}\geq 0$ corresponds to the inconclusive outcome. 
Since the source states~\eqref{psik} are linearly independent, it is possible to find a set of orthogonal states $\{ | \tilde{\Phi}_k \!\left.\right\rangle \}_{k=1}^n$ 
such that $\braket{\tilde{\Phi}_l}{\Psi_k}=\delta_{kl}$ (the tilde indicates that these states are not normalized in general). These states can be compactly written as
$
| \tilde{\Phi}_k \!\left.\right\rangle=\Omega^{-1} \ket{\Psi_k} 
$, with
$
\Omega=\sum_k \ketbrad{\Psi_k},
$
where the inverse $\Omega^{-1}$  has to be understood in the pseudoinverse sense~\cite{horn} if necessary.
Then, the POVM elements of the unambiguous measurement simply read 
$E_{l}=\gamma_{l} |\tilde{\Phi}_l\rangle\!\langle\tilde{\Phi}_l|$,  $l=1,\ldots,n $, where the parameters $0\leq \gamma_l\leq 1$ are the conditional success probabilities of identifying each source state. We will refer to $\gamma_l$ as \emph{efficiencies}~\cite{efficiencies}.  The  success probability of identifying a change point without error 
is given by $P_{s}=\frac{1}{n}\sum_{k=1}^{n}\gamma_{k}$,  and the efficiencies $\gamma_k$,  the only free parameters left to be optimized, are constrained by the condition $E_0\geq0$. 
 
The optimal efficiencies, up to a certain critical value $c^*$ of the overlap,
are~\cite{cp-2}
\begin{align}
\label{efficiencies}
\gamma_n(k) = \sum_{j=1}^n (-c)^{|k-j|} \,,\quad k=1,\ldots,n \, ,
\end{align}
where we have explicitly included the dependence on the number $n$ of particles and written $\gamma_k$ as $\gamma_n(k)$. The corresponding optimal success probability reads 
\begin{eqnarray}
\label{ps-upper}
P_{s}=\frac{1}{n}\sum_{k=1}^{n}\gamma_{n}(k)=\frac{1-c}{1+c}+ \frac{1}{n} \frac{2c\left[1-(-c)^n\right]}{(1+c)^2} \,.
\end{eqnarray}
This expression is valid in the range $0\leq c \leq c^*$, where 
$c^*\approx (\sqrt{5}-1)/2$ is determined by the equation \mbox{$\gamma_n(2)=0$}.
In the rest of the range, $c^*\leq c \leq 1$, the optimal efficiencies and success probability read~\cite{cp-2}
\begin{equation}
\label{efficiencies-p}
\gamma'_{n}(k)=\gamma_n(k)-\gamma_n(2) \frac{(-c)^{|k-2|}+(-c)^{|n-k-1|}}{1+(-c)^{n-3}}
\end{equation}%
and
\begin{equation}
\label{ps-p}
P_{s}^{\prime}=\frac{1}{n}\sum_{k=1}^{n}\gamma'_n(k)=P_{\rm s}-\frac{2}{n} \frac{\gamma^2_n(2)}{1+(-c)^{n-3}}\,,
\end{equation}
respectively.

\section{Online strategies}\label{sec:local}

The optimal solution, comprised by Eqs.~\eqref{ps-upper} and \eqref{ps-p},
in principle requires a global measurement on the whole set of $n$ particles that may be infeasible to implement in practice. 
It is therefore of interest to elucidate whether the task can be achieved with online strategies that act locally on each particle, possibly assisted by classical communication between measurements, and how does their performance compare to the optimal one. Such strategies are far easier to implement in practice, and, additionally, would allow for the detection of a change point in a stream of quantum particles as soon as it occurs.
We will show that, quite extraordinarily, there is a simple online protocol 
that performs optimally for $0\leq c\leq1/2$ and needs to store only the outcome of the last measurement at each step. 
In this overlap range, this result holds true for sequences of arbitrary length $n$.
For $c > 1/2$ the best online protocol does not attain the optimal success probability, although it is remarkably close.

A change point at position $k$ can be exactly identified by a local protocol only if there are two
successive unambiguous detections:
$\ket{0}$ at position $k-1$, followed by 
$\ket{\phi}$  at position $k$. For the end-point case $k=1$ one only requires the detection of state 
$\ket{\phi}$ at the first position, while for the last change point position, $k=n$, detecting $\ket{0}$ at position $n-1$ suffices
since it is assumed that a change point has always occurred and, hence, the state of the last particle is necessarily $\ket{\phi}$
\footnote{Notice that the no-change possibility in a string of $n$ particles is formally equivalent to a promised change point with $n+1$ particles.}. 
To  
lighten the presentation, 
from now on 
we will simply write `detection' or `detect' for `unambiguous detection' or `unambiguously detect'.

Let $\mathcal{M}_n$ be a local measurement strategy for strings of $n$ particles, where each local measurement has three possible outcomes: $0$, $\phi$, and $I$, which correspond to detecting $\ket{0}$, $\ket{\phi}$, and an inconclusive result, respectively. Let $\Theta_j$ be a particular set of outcomes of the first $j$ measurements. Then, the sequence of outcomes $(\Theta_{k-2},0_{k-1},\phi_k)$ leads to the detection of a change point at position $k$.  
The probability of a successful detection of the change point given the source state $\ket{\Psi_k}$ and a measurement strategy $\mathcal{M}_n$, that we name local efficiency for position $k$, reads
\begin{equation}\label{Dnk}
D_n(k) := \sum_{\Theta_{k-2}} {\rm Pr}[(\Theta_{k-2},0_{k-1},\phi_{k})|\Psi_k,\mathcal{M}_n] \,,
\end{equation}
and the average success probability is given by
\begin{equation}\label{ps_local-1}
P_s^{\rm L} = \frac{1}{n}\sum_{k=1}^n D_n(k) \,.
\end{equation}

We characterize next the local measurements that comprise a strategy $\mathcal{M}_n$.
An optimal measurement that unambiguously discriminates between two states $\ket{0}$ and $\ket{\phi}$ that are assumed to occur with prior probabilities
$\eta_{0}$ and $\eta_{\phi}$, respectively, 
  succeeds with conditional probability $1-c\sqrt{\eta_\phi/\eta_0}$ if the state was $\ket{0}$ and $1-c\sqrt{\eta_0/\eta_\phi}$ if it was $\ket{\phi}$ ~\cite{cp-2}.
  Therefore each local measurement is determined by a  strength parameter $x:=\sqrt{\eta_\phi/\eta_0}$ that specifies its bias towards detecting $\ket{0}$ or $\ket{\phi}$. In terms of $x$ and $c$, the local conditional probabilities ${\rm Pr(outcome|state)}$ read ${\rm Pr}(0|0)=1-c x$, ${\rm Pr}(I|0)=c x$, ${\rm Pr}(\phi|\phi)=1-c/x$, and ${\rm Pr}(I|\phi)=c/x$. Obviously, ${\rm Pr}(0|\phi)={\rm Pr}(\phi|0)=0$. The positivity of these probabilities  
  bounds the strength parameter to the interval $c\leq x\leq 1/c$.
The extreme value $x=c$  ($x=1/c$) corresponds to an effective two-outcome measurement that either detects $\ket{0}$ ($\ket{\phi}$) or yields an inconclusive answer, and any other intermediate value of $x$ represents 
a three-outcome measurement. An optimal local measurement strategy $\mathcal{M}_n$ is  
a sequence of $n-1$ unambiguous measurements 
that maximizes Eq.~\eqref{ps_local-1}.

We address the problem of finding the optimal  $\mathcal{M}_n$  
by considering general
adaptive strategies that take into account the information learned in previous measurements. We introduce this feature by letting the strength of the measurement over particle $j$ generically depend on all past outcomes $\r_{j-1}=\{r_1,\ldots,r_{j-1}\}$, that is, 
$x(j;\r_{j-1})$.
Note that 
$\r_{j-1}$
cannot contain any outcome $\phi$, as the procedure stops after obtaining the first $\phi$.
Thus, $\r_{j-1}$ is a binary string of $0$'s and $I$'s.  
This is the most general one-way local-operations-and-classical-communication (LOCC) protocol that one can devise~\cite{locc-rmt}.
Optimizing LOCC protocols is in general unfeasible, since the number of parameters grows exponentially with $n$. 
However, for the problem at hand, this number is effectively reduced to $n-1$ and thus the optimization can be tackled efficiently. This 
exponential 
reduction is a direct consequence of the logic behind unambiguous measurements: after obtaining an outcome $0$ at position $j$, one knows for a fact that all particles of the string up to the $j$th position were in the state $\ket{0}$, therefore any information that previous outcomes may provide is superseded.
Further, if the outcome of the measurement over the $j$th particle is $I$, the following optimal measurement strength is fixed to detect only the state $\ket{0}$, since 
the sequence of outcomes $I\phi$ 
irremediably implies the failure of the protocol. These observations are condensed in the equations
$x(j;r_{j-1}=0)=:x(j)$, $x(j;r_{j-1}=I)=c$, 
hence the free parameters of a general adaptive strategy $\mathcal{M}_n$ 
is just the set of strengths 
$\{x(j)\}_{j=1}^{n-1}$ 
of measurements that are preceded by an outcome $0$.

To gain intuition on the general solution, we first show the explicit construction of the optimal strategy for \mbox{$n=4$}. The conditional detection probabilities are
\begin{align}
 D_4(1) =& \,1-\frac{c}{x(1)} \,,\label{d-41}\\
 D_4(2) =& \,\left[1-c\,x(1)\right]\left[1-\frac{c}{x(2)}\right] \,,\label{d-42}\\
 D_4(3) =& \,\left[1-c\,x(1)\right]\left[1-c\,x(2)\right]\left[1-\frac{c}{x(3)}\right] +\nonumber\\
& \,c\,x(1) (1-c^2)\left[1-\frac{c}{x(3)}\right]\,,\label{d-43}\\
 D_4(4) =& \,\left[1-c\,x(1)\right]\left[1-c\,x(2)\right]\left[1-c\,x(3)\right] \nonumber\\
 & + c\,x(1) (1-c^2)\left[1-c\,x(3)\right] \nonumber\\
& +\left[1-c\,x(1)\right]c\,x(2) (1-c^2) + c\,x(1) c^2 (1-c^2) \,.\label{d-44}
\end{align}
Each summand in $D_4(k)$ corresponds to the probability of a string of outcomes leading to detection of the change point at position $k$. For instance, $D_4(4)$ comprises the strings $000$, $I00$, $0I0$, and $II0$. The maximization of Eq.~\eqref{ps_local-1} leads to the optimal strengths
\begin{equation}
\label{x-4}
x(1)=\frac{1}{1-c+c^2}, \; x(2)=\frac{1}{1-c}, \; x(3)=1 \,.
\end{equation}
The first key observation is that the optimal local efficiencies match the optimal global efficiencies for each change point, and, therefore, $P_s^{\rm L}=P_s$. Indeed, inserting Eq.~\eqref{x-4} into Eqs.~\eqref{d-41} to \eqref{d-44}, one obtains $D_4(k) = \gamma_4(k)$ for $k=1,\ldots,4$, where $\gamma_4(k)$ is given in Eq.~\eqref{efficiencies}. The second key observation is that this solution only holds for $c\leq 1/2$, since outside this range 
$x(2)>1/c$
and hence it does not yield a valid measurement. Further, the optimal value 
$x(3)=1$ 
can be easily understood: conditioned to having obtained $r_2=0$, the probability of the third particle being in the state $\ket{0}$ or  $\ket{\phi }$ is $1/2$,
hence the optimal choice is a symmetric measurement.
We will see that these features remain valid in the general case.

\subsection{Optimal online protocol}

Let us now present the solution for the optimal strength parameters and detection probabilities for arbitrary $n$.  It is convenient to write the explicit dependence on the total number of particles, i.e., $x(j)$ 
as $x_n(j)$. 
As discussed before, obtaining an outcome $0$ at position $j-1$ discards all hypotheses with a change point at position $k\leq j-1$, effectively resetting the problem to one with a change point in a string of $n-j+1$ particles. Hence, we have that $x_n(j)=x_{n-j+1}(1)$ holds for optimal strengths. Now, we follow the intuition from the $n=4$ problem that, in case there is no performance gap between the optimal global and local strategies, the global and local efficiencies should match one by one. This leads us to the equation $D_m(1)=1-c/x_m(1)=\gamma_m(1)$ [cf. Eq.~\eqref{d-41}]. Using the explicit value of $\gamma_{n-j+1}(1)$ from Eq.~\eqref{efficiencies}, we obtain
\begin{equation}\label{xn-k}
x_n(j)=x_{n-j+1}(1)=\frac{1+c}{1-(-c)^{n-j}} \,,\quad j=1,\ldots,n-1\,.
\end{equation}
Note that this formula reduces to Eq.~\eqref{x-4} for $n=4$, and that it is a solution for the set of equations $\{D_n(k)=\gamma_n(k):k=1,\ldots,n-1\}$.  In Appendix~\ref{app-A} we provide a proof by induction of Eq.~\ref{xn-k}. We also note that \mbox{$x_n(n-2)=1/(1-c)$} is the first strength to saturate at the extreme value $1/c$ with increasing $c$, hence this general solution is still only valid up to $c= 1/2$.
In summary, for overlaps $0\leq c \leq 1/2$, the optimal online strategy consists in a sequence of unambiguous measurements of strengths $x_n(j)$ given by Eq.~\eqref{xn-k} if the outcome of the measurement on the previous particle is $0$, and fixed strengths $c$ if the previous outcome is $I$. 
This online protocol attains the performance of the optimal (global) strategy, given by Eq.~\eqref{ps-upper}.
  
\subsection{Beyond $c>1/2$}

We now analyze the optimal local strategy for $c>1/2$. It is clear that local strategies cannot reach optimal performance in this range of overlaps, as this would require the expressions of the strengths \eqref{xn-k} remain valid beyond their upper limit $1/c$. The optimization of local protocols  
 is much more constrained than that of a global strategy and, hence, a smaller feasibility region is to be expected.
 As $c$ increases, there is a progressive saturation of the strengths, starting with $x_n(n-2)$. The exact saturation point for each strength can be computed from the techniques shown in Appendix~\ref{app-B}, as well as
the point $c_S\approx 0.69$ where all strengths but the last one [always fixed to be $x_{n}(n-1)=1$] are saturated at $x_n(k)=1/c$.
Beyond $c_S$, the optimal online strategy is a sequence of two-outcome unambiguous measurements, aiming at the detection of $0$ ($\phi$) if the previous outcome was $I$ ($0$). 

The exact expressions for the optimal local protocol in the intermediate region $1/2<c<c_S$ are rather impractical. Instead, we provide a simpler local protocol that we prove to be optimal for large $n$. By doing so, we also discover that online protocols can still attain optimal global performance beyond $c=1/2$ and up to $ c=c^*$,
precisely the value that divides the two regimes in the global approach. 
We consider the  simple 
local strategy with constant strengths 
$x_n(k)=x$  after a $0$ outcome and, of course, a strength $c$ after an $I$ outcome.
In Appendix~\ref{app-C} we show that the success probability of such strategy for large $n$ reads
\begin{equation} 
\label{FL-1}
P_s \simeq \frac{1-c^2}{1+ c x -c^2}\left( 1-\frac{c}{x} \right)\,,
\end{equation} 
which is maximal for $x=1+c$. Note that we could 
have anticipated this result, as it corresponds to the approximation of Eq.~\eqref{xn-k} for large $n$. The maximal success probability for local strategies with constant strengths then reads
\begin{equation} 
\label{FL-2}
P_s^{\rm FL} \simeq \frac{1-c}{1+ c}\,,
\end{equation} 
which coincides with the optimal asymptotic value in Eqs.~\eqref{ps-upper} and \eqref{ps-p} (the superscript FL stands for fixed local). 
The choice  $x=1+c$  yields a valid a measurement up to \mbox{$c^*\approx 0.61$}, a value that is determined by the saturation condition $1+c=1/c$.

\begin{figure}[ht]
	\includegraphics[scale=0.33]{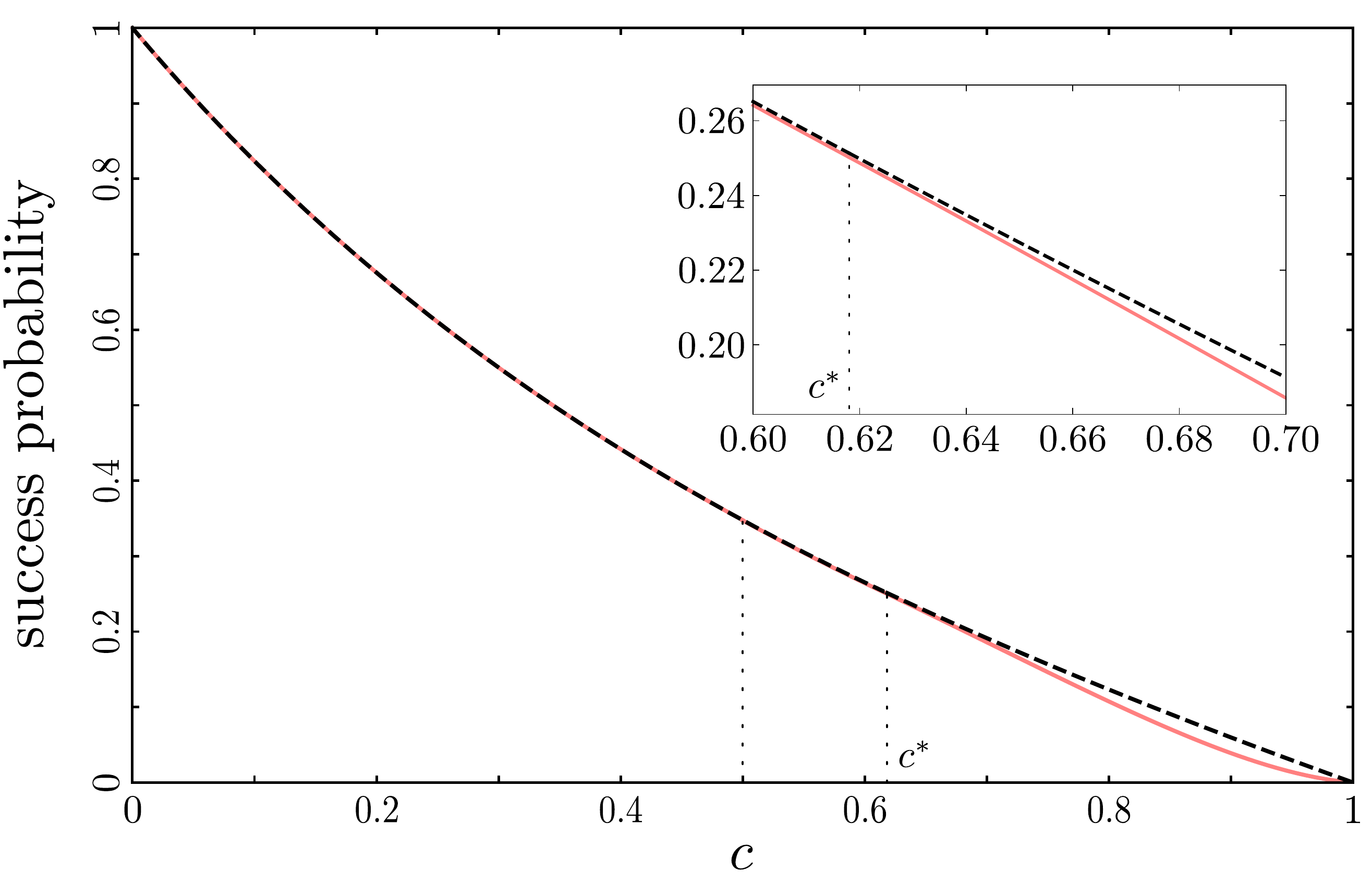}
	\caption{Probability of exact identification of a change point as a function of $c=|\!\braket{0}{\phi}\!|$ for a string of $n=31$ particles. The black dashed line is the optimal success probability 
when global strategies are considered, given by the piecewise function composed by Eqs.~\eqref{ps-upper} and \eqref{ps-p}. The pink solid line corresponds to the success probability of different online detection strategies, depending on the value of $c$. In the interval $0\leq c\leq 1/2$, the online strategy characterized by the strengths in Eq.~\eqref{xn-k} matches exactly the optimal performance. Beyond $c=1/2$ the FL strategy is optimal only asymptotically and up to $c=c^*$, although the difference for $n=31$ of around $\sim 0.1\%$ is hardly appreciated. For $c>c^*$ the success probability of the SL strategy starts to deviate from optimality and shows a finite gap even in the asymptotic limit. The inset plot highlights this regime transition of online strategies around $c^*$.}\label{fig1} 
\end{figure}

For $c>c^*$, the constant strengths saturate to $x=1/c$ and Eq.~\eqref{FL-1} reads 
\begin{equation} 
P_s^{\rm SL} \simeq \frac{(1-c^2)^2}{2-c^2}\,,
\end{equation} 
where SL stands for saturated local. The success probability deviates from the optimal value given in Eq.~\eqref{FL-2}, but the difference is no larger than 2.2\% in the worst case~\footnote{Note that the SL strategy is, strictly speaking, suboptimal among online strategies in the interval $0.61 \approx c^* \le c \le c_S\approx 0.69$. Although the optimal online strategy can always be constructed following the techniques in Appendix~\ref{app-B}, the difference in performance with respect to the SL strategy in this small range of $c$ is inappreciable}. 
The closeness of online protocols to optimal performance is patent in Fig.~\ref{fig1}, where we represent the average success probability of the best online strategy in each overlap regime together with the optimal one for a string of length $n=31$.

\section{Conclusions}\label{sec:conclusions}

Let us conclude by reviewing our results and its implications. 
In this work, we have derived the optimal local protocol that unambiguously detects a quantum change point. We have shown that it attains exactly the performance of an optimal global protocol for arbitrary string lengths and for values of the overlap $c$ below 1/2. Our results provide not only one of the few non-trivial examples of state identification tasks where the optimal protocol can be found, but 
also the first
instance with 
arbitrarily many
 hypotheses where one-way LOCC measurements match optimal performance
 (see Refs.~\cite{bayesian,Zhengfeng2005} and \cite{Chefles2001,Nakahira2018} for such instances for binary and ternary discrimination, respectively).
 Besides this remarkable feature, the LOCC protocol  has several attractive  aspects:  (i) it is an online protocol, i.e., in case the change point is detected, it is as soon as it appears; (ii) no quantum memories are required;  (iii) the necessary measurements are all local and, hence,  easy to implement experimentally; 
and (iv) the memory required for the adaptive selection of subsequent measurements amounts to just one bit (encoding whether the previous outcome was conclusive or not),
which may benefit the stability and robustness of an experimental setup.  

We have also analyzed how above $c=1/2$ the problem becomes too constrained for any online strategy to attain global optimality for strings of arbitrary finite length, although the performance gap is very small. Despite this, for large $n$, we have shown that optimal global performance can still be attained for overlaps up to $c^*\approx 0.61$ by an online fixed-strength strategy.
Beyond $c^*$, the best local protocol essentially consists in a sequence of two-outcome measurements that detect just one of the local states: either $\ket{0}$ or $\ket{\phi}$. We have shown that such protocol deviates from the optimal performance only by 2.2\% in the worst case.

Finally, it is worth mentioning that our results are amenable to experimental realization with current technology,  as 
the experimental implementation of the necessary unambiguous measurements has already been demonstrated
in optical platforms~\cite{exp,exp-coherent,exp-random,exp-random-2}.


\acknowledgments
This research was supported by the Spanish MINECO
through contract FIS2016-80681-P, the ERC Consolidator Grant 683107/TempoQ and the DFG. EMV acknowledges financial support from CONACYT.

\bibliographystyle{unsrt}

\appendix
\section{Proof of Eq.~\eqref{xn-k}}\label{app-A}
In this section we provide a proof by induction of the optimal form of the strengths $x_n(k)$, given by Eq.~\eqref{xn-k}. 
This optimal form is
\begin{equation}
\label{supp-xn-k}
x_{n}(k)=\frac{1+c}{1-(-c)^{n-k}}\,.
\end{equation}
Let us first establish some additional notation. Recalling Eq.~\eqref{Dnk}, 
given a local strategy $\mathcal{M}_n$, the efficiency of detection of the change point at position $k$ reads
\begin{equation}\label{dnk_supp}
\begin{split}
D_n(k) &= \sum_{\Theta_{k-2}} {\rm Pr}[(\Theta_{k-2},0_{k-1},\phi_{k})|\Psi_k,\mathcal{M}_n] \\
&=: \P(\Sigma_{k-2},0_{k-1},\phi_k) \,,
\end{split}
\end{equation}
where the sum runs over the $2^{k-2}$ sets that only contain outcomes $0$ and $I$, and $\Sigma_j$ denotes all such sets of $j$ outcomes.
The argument of $\P$ is always to be understood as an ordered, consecutive sequence of outcomes, so we will omit the position subscripts when no confusion arises.
As argued in the main text, the optimal local protocol can be obtained by equating each local efficiency to the corresponding global one, i.e., $D_n(k)=\gamma_n(k)$, and solving the resulting system of equations. Recall that the optimal global detection efficiencies read
\begin{align}
\label{efficiencies-supp}
\gamma_n(k) = \sum_{j=1}^n (-c)^{|k-j|} = \frac{1-c-(-c)^k-(-c)^{n-k+1}}{1+c}
\end{align}
for $k=1,\ldots,n$.
We first observe that, for $k<n$, these equations 
read
\begin{equation}
\P(\Sigma_{k-2},0,\phi) =  \P(\Sigma_{k-2},0)\left[1-\frac{c}{x_n(k)}\right]= \gamma_n(k)\,.
    \label{ec:genstruct}
\end{equation}
We also have that
\begin{multline}
    \P(\Sigma_{k-1},0) = \P(\Sigma_{k-2},0)[1-c\,x_n(k)]\\
    +\P(\Sigma_{k-2},I)(1-c^2)\,,
    \label{ec:probstruct}
\end{multline}
and recall that $\P(\Sigma_{k-2},I)=1-\P(\Sigma_{k-2},0)$.
Then, using Eqs.~\eqref{ec:genstruct} and \eqref{ec:probstruct} we get the relation
\begin{equation}
    x_n(k+1)=c\left[1-\frac{\gamma_n(k+1)}{(1-c^2)-c\,\gamma_n(k)\,x_n(k)}\right]^{-1}\,.
    \label{ec:recursivexk}
\end{equation}
The first strength is immediate to derive from the equation $D_n(1)=1-c/x_n(1)=\gamma_n(1)$:
\begin{align}
    \label{ec:inverse}
    x_n(1)  =\frac{c}{1-\gamma_n(1)}=\frac{1+c}{1-(-c)^{n-1}}\,,
\end{align}
where we have used Eq.~\eqref{efficiencies-supp} for $k=1$. 
Using Eqs.~\eqref{ec:recursivexk} and \eqref{efficiencies-supp}, we arrive by induction at the formula for the optimal strengths Eq.~\eqref{supp-xn-k}.

The attentive reader should have noticed that 
the system of equations $D_n(n)=\gamma_n(n)$ for $k=1,\ldots,n$ is over-constrained: there are $n$ equations but $n-1$ unknowns $x_n(k)$. The first $n-1$ equations determine univocally all the unknowns, and the last equation, $D_n(n)=\gamma_n(n)$, should be automatically satisfied. This could seem at first sight a rather non-trivial requirement as $D_n(n)$ contains $2^{n-2}$ summands (such is the size of the set $\Sigma_{n-2}$), but the proof is quite straightforward. We recall that $\gamma_n(n-1)=D_n(n-1)=\P(\Sigma_{n-3},0_{n-2},\phi_{n-1})=\P(\Sigma_{n-3},0_{n-2},0_{n-1})=(1-c)\P(\Sigma_{n-3},0_{n-2})$, because the last strength takes the symmetric value $x_n(n-1)=1$. Then,
\begin{align}
D_n(n) & =\P(\Sigma_{n-3},0,0)+\P(\Sigma_{n-3},{I},0) \nonumber \\
            & =(1-c) \P(\Sigma_{n-3},0)+(1-c^2) \P(\Sigma_{n-3},{I}) \nonumber \\ 
             & =(1-c)\P(\Sigma_{n-3},0)+(1-c^2)[1-\P(\Sigma_{n-3},0)] \nonumber \\
             &=(1-c^2) -(1-c)\P(\Sigma_{n-3},0) \nonumber \\
             &=(1-c^2) -\gamma_n(n-1) =\gamma_n(n),
\end{align}
where the last equality  can be easily checked from Eq.~\eqref{efficiencies-supp}.

\section{Construction of optimal local strategies}\label{app-B}

Here we show a general method to construct an optimal set of strengths for any given $n$. This method is particularly useful in the range of overlaps $1/2 < c \le c_S\approx 0.69$, where a mixture of saturated and unsaturated strengths coexist.
Given an arbitrary local strategy determined by the set of strengths $\{x_n(k)\}_{k=1}^{n-1}$, we write the maximization conditions $\partial P_s^{\rm L}/ \partial x_n(k)=0$, 
where $P_s^{\rm L}=(1/n)\sum_{k=1}^n D_n(k)$ and $D_n(k)$ is given by Eq.~\eqref{dnk_supp}.
Starting from the last strength, we note that all the terms  of $P_s^{\rm L}$ that depend on  $x_{n}(n-1)$ can be written as 
\begin{equation}\label{eq:B1}
\begin{split}
\P(\Sigma_{n-3},0_{n-2})\left[{\rm Pr}(\phi_{n-1}|0_{n-2})+ {\rm Pr}(0_{n-1}|0_{n-2}) \right]\\
=\P(\Sigma_{n-3},0_{n-2})\left[1-\frac{c}{x_{n}(n-1)}+1-c x_{n}(n-1)\right]\,,
\end{split}
\end{equation}
where the last factor takes the same form for any value of $n$. The two probabilities inside the brackets in the first line of Eq.~\eqref{eq:B1} are, respectively, the probabilities of obtaining outcomes $\phi$ and $0$ at position $n-1$ conditioned on an outcome $0$ at position $n-2$. Note that both events successfully identify change points at positions $n-1$ and $n$, respectively.
The maximization of Eq.~\eqref{eq:B1} yields $x_n(n-1)=1$. This value intuitively makes sense, since measuring the state of the particle at position $n-1$ means that only two equally likely possible change points remain, either at position $n-1$ or at position $n$. In such binary identification case, it is clear that a balanced measurement is optimal.

Next we  write all the terms of $P_s^{\rm L}$ that depend on $x_{n}(n-2)$, and substitute the value $x_n(n-1)=1$. We obtain
\begin{equation}
\label{P-n-2}
\begin{split}
\P(\Sigma_{n-4},0_{n-3}) \Big\{&1-\frac{c}{x_{n}(n-2)}+2(1-c)[1-c x_{n}(n-2)]\\
&+(1-c^2) c x_{n}(n-2)\Big\}.
\end{split}
\end{equation}
Again, the term in brackets is the same for any $n$, and it determines the optimal value $x_{n}(n-2)=1/(1-c)$. 
One can proceed recursively, and realize that the optimal value of $x_{n}(n-j)$ satisfies  $x_{n}(n-j)=x_{n+1}(n+1-j)$. This observation provides an alternative proof
that $x_n(k)=x_{n-1}(k-1)$ is not only a feature of the optimal unsaturated strengths [cf. Eq.~\eqref{supp-xn-k}], but also holds for optimal strengths in general, even when the extremal conditions $\partial P_s^{\rm L}/ \partial x_{n}(k)=0$ are not satisfied (which happens when the feasibility constraint $ x_n(k)\leq 1/c$ is hit). In this situation, one substitutes the strengths for its extremal value and carries on with the maximization of the next strength. 

For the subsequent strengths it is convenient to use the notation $\Sigma^{k_2}_{k_1}$ to denote all the possible strings of outcomes between particle $k_2$ and $k_1$.
For  $x_n(n-3)$,  we have $\Sigma^{n-3}_{n-1}=\{II0, I00,000,I0\phi, 00\phi\}$, and we have to consider the conditional probability 
 ${\rm Pr}( \Sigma^{n-3}_{n-1}|0_{n-4})$ with $x_n(n-2)=1/c$ and $x_n(n-1)=1$.  Solving $\partial {\rm Pr}( \Sigma^{n-3}_{n-1}|0_{n-4})/\partial x_n(n-3) =0$,  we obtain
\begin{equation}
\label{xp-1}
x_n(n-3)=\frac{1}{\sqrt{c(2-c)(1-c^2)}}\,.
\end{equation}
The saturation condition $x_n(n-3)=1/c$ has the solution $c=:c_S\approx 0.69$. Checking for several values of $n$, one sees that $x_n(n-3)$ is always the last strength that reaches the saturation point. 
This is in accordance with the intuition that the smallest unsaturated strength [of the form in Eq.~\eqref{supp-xn-k}] should be the last one to reach $1/c$.
Then, $c_S$ corresponds to the total saturation point, defined as the point beyond which all strengths are saturated [naturally with the exception $x_n(n-1)=1$].
 
For the following strengths $x_n(k)$ for $k=n-4,n-5,\ldots$, one proceeds by recursively maximizing ${\rm Pr}( \Sigma^{n-k}_{n-1}|0_{n-k-1})$, taking into account if the
saturation condition is hit for any of the strengths. Notice that only one variable, $x_n(k)$, is maximized at each step, because the strengths $x_n(k+1),x_n(k+2),\ldots,x_n(n-1)$ have been already fixed at their optimized value obtained in previous optimization steps.
\section{The fixed local strategy}\label{app-C}

Here we include the calculation of the success probability for the fixed local (FL) strategy, and explicitly show that, in its range of validity, it is asymptotically optimal. Let us first rename $\P(\Sigma_{k-1},{I})=:G(k)$, and note that $\P(\Sigma_{k-1},0)=1-G(k)$. Then, for a generic local strategy with fixed strengths $x_n(k)=:x$, we can write
\begin{align}
G(k+1)=& \P(\Sigma_k,{I})=\P(\Sigma_{k-1},{I},{I})+\P(\Sigma_{k-1},0,{I})\nonumber \\
           =& c^2\P(\Sigma_{k-1},{I})+c\,x\,\P(\Sigma_{k-1},0)\nonumber  \\
           =& c^2 G(k)+c\,x\,[1-G(k)] \nonumber \\
           =& c\,x-(c\,x-c^2)G(k)\,, \label{g-recursion}
\end{align}
where we have used that $\P({I},{I})=c^2\P({I})$ (recall that, after an outcome $I$, we always apply an extreme two-outcome unambiguous measurement completely biased towards detection of the local state $\ket{0}$), and $\P(0,{I})=c\,x\,\P(0)$. 
The recursion relation~\eqref{g-recursion} can be readily solved using the initial condition $G(1)=c\,x$ to give
\begin{align}
\label{g-function}
G(k)=c\,x \frac{1-(c^2-c\,x)^k}{1+c\,x -c^2}\,.
\end{align}
We can relate the local detection efficiencies $D_n(k)$ to the function $G(k)$ by looking at Eq.~\eqref{dnk_supp}. We obtain
\begin{multline}\label{dnk-g}
D_n(k)=G(k-2)(1-c^2)\left(1-\frac{c}{x}\right)\\
+[1-G(k-2)](1-c\,x)\left(1-\frac{c}{x}\right)\,,
\end{multline}
which is valid for $k=1,\ldots,n-2$. Note that, while the original definition $G(k)=\P(\Sigma_{k-1},I)$ does not hold physical meaning in the cases $G(-1)$ and $G(0)$, using the functional expression \eqref{g-function} in Eq.~\eqref{dnk-g} we recover the correct local efficiencies for the first and second positions, namely
$D_n(1)=1-c/x$ and $D_n(2)=(1-c\,x)(1-c/x)$.
The last two local efficiencies have a slightly different expression and cannot be recovered from Eq.~\eqref{dnk-g}. This is so because the last strength is always fixed to $x_{n-1}=1$ conditioned to having obtained $r_{n-2}=0$ as a previous outcome, as argued in the main text. In addition, recall that the $n$th particle is in the state $\ket{\phi}$ by definition and hence there is no need to measure it.
Taking this into account, the success probability for the FL strategy can be written as
\begin{align}
\label{ps-f}
P_s^{\mathrm{FL}} &=\frac{1}{n}\left\{\sum_{k=1}^{n-2} G(k-2) (1-c^2)\left(1-\frac{c}{x}\right)\right.\nonumber \\
              +& \left.\left[1- G(k-2)   \right] (1-c x)\left(1-\frac{c}{x}\right) \right.\nonumber \\
              + &\left.\left\{G(n-3)(1-c^2) + [1-G(n-3)](1-c\,x)\right\}(1-c) \right.\nonumber\\
              + &\left.G(n-2)(1-c^2)+[1-G(n-2)](1-c)\phantom{\sum_k^n}\hspace{-0.5cm}\right\} \nonumber\\
              &=\frac{1}{n}\left\{\sum_{k=1}^{n-2} G(k-2) (1-c^2)\left(1-\frac{c}{x}\right)\right.\nonumber\\
              &\phantom{XXX}+\left[1- G(k-2)   \right] (1-c x)\left(1-\frac{c}{x}\right) \nonumber \\
              &\phantom{XXX}+\left. (1-c)\left[2-(1-c)G(n-2)\right]\phantom{\sum_k^n}\hspace{-0.5cm}\right\}\,.
\end{align}

For large $n$, the leading order of the success probability is 
\begin{align}
\label{ps-f-as}
P_s^{\mathrm{FL}}  & \simeq \frac{c x}{1+c x -c^2} (1-c^2)\left(1-\frac{c}{x}\right)             \nonumber\\
&\quad+ \frac{1-c^2}{1+c x -c^2} (1-c x)\left(1-\frac{c}{x}\right)\nonumber\\
&=\frac{1}{1+c\,x-c^2}-\frac{c}{x}\,,
\end{align}
which just amounts to neglect the exponential terms in  Eq.~\eqref{g-function} and the slightly different last term in Eq.~\eqref{ps-f}. Note that this asymptotic limit of the success probability would remain invariant if we would choose the same fixed strength $x_{n-1}=x$ for the last measurement too, as opposed to the slightly better choice of a symmetric strength $x_{n-1}=1$. Eq.~\eqref{ps-f-as} can be easily  maximized to obtain
\begin{equation}
\label{ps-f-as-2}
x_{\mathrm max}=1+c \ \ \ \to \ \ P_s^{\mathrm{FL}}   \simeq \frac{1-c}{1+c}\,.
\end{equation}
Hence, as anticipated, we obtain that a protocol that measures a particle with a local measurement of fixed strength $x=1+c$ if the previous outcome was 0, and $x=c$ if the previous outcome was inconclusive, is asymptotically optimal up to a threshold value of the overlap $c^*\approx 0.61$. This threshold is the solution of the boundary constraint on the fixed strength $x$, i.e., $1+c=1/c$.

\section{Success probability of the Saturated Local strategy}\label{app-D}

The saturated local (SL) strategy is defined by the fixed strengths $x_n(k)=1/c$, at the boundary of their physicality interval. The corresponding local efficiencies, $D_n(k)$, up to $k=n-2$ read
\begin{equation}
D_n(1)=(1-c^2), \ \ D_n(2)=0\,, 
\end{equation}
and
\begin{align}
\label{dn-k_supp}
D_n(k)=(1-c^2)^2 F(k-2), \ \ k=3,\ldots, n-2 ,
\end{align}
where the function $F(k)$ can be directly read off of Eq.~\eqref{g-function} particularizing for $x=1/c$. The exact expression for the success probability and the leading order term in the asymptotic regime of large $n$ are derived likewise from Eqs.~\eqref{ps-f} and \eqref{ps-f-as}. The latter reads
\begin{equation}\label{pl-asymptotic_supp}
P_s^{\rm SL} \simeq \frac{(1-c^2)^2}{2-c^2}.
\end{equation}
Notice that this value is smaller than the leading term $(1-c)/(1+c)$ of the optimal success probability [cf. Eqs.~\eqref{ps-upper}, \eqref{ps-p}, and Eq.~\eqref{ps-f-as-2}]. The difference is however very small, with a maximal value of $0.022$ at $c\approx 0.89$.  The asymptotic success probability for the SL strategy, Eq.~\eqref{pl-asymptotic_supp}, equals the optimal value precisely at $c^*=(\sqrt{5}-1)/2$, below which the FL strategy is asymptotically optimal, as discussed in the main text.

\end{document}